\documentclass[11pt,a4paper]{article}
\usepackage[margin=1in]{geometry}
\usepackage{graphicx}
\usepackage[hidelinks]{hyperref}
\usepackage{authblk}
\usepackage{caption}
\usepackage{subcaption}
\usepackage{float}
\usepackage{hyperref}

\usepackage[left]{lineno}
\usepackage{times}

\title{The CMS ME0 Upgrade: Enhancing Forward Muon Reconstruction at the HL-LHC}
\author{Anureet Kaur$^{1}$ for the CMS Collaboration\\
$^{1}$Panjab University, Chandigarh, India}
\date{\vspace{-1em}}

\begin{document}
\maketitle
\vspace{-1em}
\begin{center}
Presented at the 32nd International Symposium on Lepton Photon Interactions at High Energies,\\
Madison, Wisconsin, USA, August 25--29, 2025.
\end{center}

\begin{abstract}
The CMS muon system is undergoing substantial upgrades to meet the challenges of the High-Luminosity LHC (HL-LHC), including the installation of the new Muon Endcap 0 (ME0) detector. Large-scale production started in 2024. ME0 is a six-layer station designed to extend pseudo-rapidity coverage to $|\eta| = 2.8$ from the previous maximum of $|\eta| = 2.4$, enhancing sensitivity to forward physics processes. Each endcap will host 18 ME0 stacks, with each stack comprising six triple-layer gas electron multiplier (GEM) chambers. The system adds up to six additional hits per track, which significantly improves muon identification, spatial resolution, and robust track reconstruction at the first trigger level. Chamber production and quality control across multiple international sites ensure scalability and timely delivery. The ME0 design incorporates lessons learned from earlier GEM deployments, with improvements in electronics robustness, grounding, and segmentation to withstand high background rates and minimize damage from discharges. This contribution provides a comprehensive overview of the ME0 detector concept, assembly strategy, quality assurance procedures, current production status, and its pivotal role in strengthening CMS muon reconstruction during HL-LHC operations.
\end{abstract}

\section{Introduction}
The HL-LHC aims to increase the instantaneous luminosity by a factor of five to seven, reaching $(5{-}7)\times10^{34}$~cm$^{-2}$s$^{-1}$. Although the Compact Muon Solenoid (CMS) experiment~\cite{CMS2008} has proven highly efficient, its forward muon region remains sensitive to radiation and high particle rates. To sustain performance under such conditions, the CMS muon system is being upgraded with new GEM detectors~\cite{CMSMuonTDR,CMSGEMTDR}: GE1/1, GE2/1, and ME0.

\begin{figure}[htbp]
\centering
\includegraphics[width=0.8\textwidth]{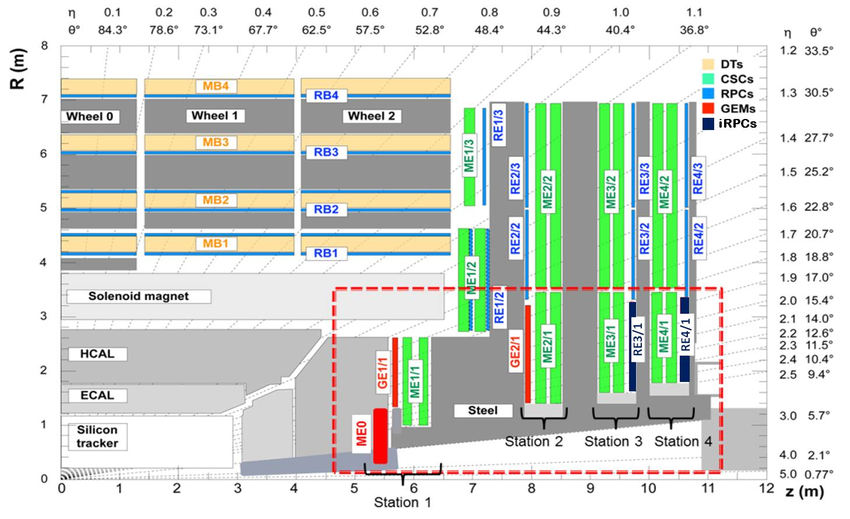}
\caption{Layout of the CMS muon endcap region showing GE1/1, GE2/1, and ME0 stations, extending muon coverage up to $|\eta| \approx 2.8$.}
\label{fig:CMS_muon_endcap}
\end{figure}

In the upgraded endcap region, the ME0 station will be the first muon detection layer encountered by a track originating from the interaction point. It will be positioned just behind the High-Granularity Calorimeter (HGCal) and will extend the muon coverage to $|\eta| \approx 2.8$, thereby improving tracking performance, trigger efficiency, and background rejection in the forward region. Figure~\ref{fig:CMS_muon_endcap} shows the location of the ME0 station within the CMS detector.

\section{The ME0 GEM Detector}

\subsection{ME0 Design and Performance}
The ``ME0'' designation refers to the Muon Endcap (ME) system, with the ``0'' indicating that this new muon station is positioned in front of the existing CMS endcap muon detectors. The ME0 detector is a critical addition to the CMS upgrade program, designed to ensure efficient muon reconstruction and triggering under the extreme conditions of the HL-LHC. Its implementation mitigates the effects of aging detector components and reduced spatial resolution in the high-radiation forward region.

The ME0 detector is composed of trapezoidal stacks, each comprising six triple-GEM modules, arranged in a planar ring on each endcap. Each ring, centered around the beamline, has an inner radius of approximately 0.6~m and an outer radius of 1.5~m. Each ring contains 18~stacks and each stack spans an azimuthal angle of $\Delta\phi \approx 20^{\circ}$, extending the muon acceptance beyond the previous limit of $|\eta|=2.4$.

The performance goals for ME0 are particularly demanding due to the intense background and radiation environment near the beamline. The detector is designed to achieve:
\begin{itemize}
\item A single-module efficiency above 97\% and a six-layer stack efficiency of approximately 98.8\%;
\item Operation under particle rates up to 150~kHz/cm$^2$;
\item Timing resolution between 8--10~ns for accurate bunch-crossing identification;
\item Gain uniformity within 15\% across chambers and long-term gain stability after accumulating 840~mC/cm$^2$;
\item Radiation tolerance up to 7.9~C/cm$^2$, ensuring reliable operation over the HL-LHC lifetime.
\end{itemize}

The trapezoidal GEM foils are segmented into 40~independent sectors, each smaller than 100~cm$^2$, to minimize the energy released during discharges and to insure uniform rate capability along the $\eta$ direction. The system provides up to six additional hits per muon track, improving pattern recognition, momentum resolution, and muon identification at the Level-1 (L1) trigger.

\subsection{ME0 Electronics}

The ME0 readout system is engineered to handle high hit rates with minimal dead time and data loss. Its design follows the well-established GEM readout architecture developed for GE1/1 and GE2/1, ensuring compatibility and reliability. The main components include:
\begin{itemize}
\item \textbf{VFAT3 (Very Forward ATLAS and TOTEM v3)} ASICs, which provide digital readout for 128~strips each and ensure low-noise signal processing with zero suppression;
\item \textbf{GEM Electronics Board (GEB)}, a 1~mm  thick printed circuit board that routes signals and provides shielding between the front-end chips and the back-end optical interface;
\item \textbf{OptoHybrid (OH)} boards, which aggregate data from six VFAT3 chips and transmit it via radiation-tolerant \textit{lpGBT} transceivers to the back-end DAQ system;
\item \textbf{bPOL DC--DC converters}, which regulate and distribute stable low-voltage power to the front-end electronics.
\end{itemize}

All components go through extensive laboratory and test-beam validation to confirm their performance under high-rate conditions. The modular structure of the ME0 electronics ensures ease of replacement.

\section{Production and Quality Control}
\begin{figure}[htbp]
\centering
\begin{minipage}[b]{0.48\textwidth}
\centering
\includegraphics[width=\textwidth]{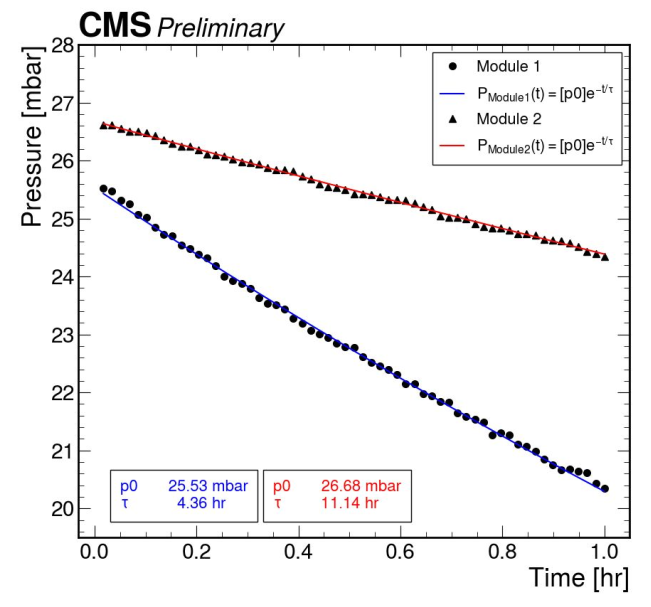}
\end{minipage}
\hfill
\begin{minipage}[b]{0.45\textwidth}
\centering
\includegraphics[width=\textwidth]{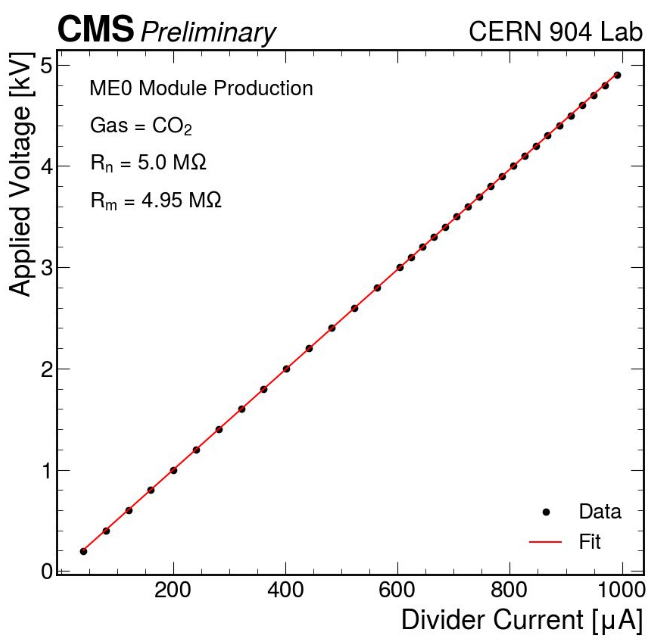}
\end{minipage}
\caption{Quality control results for ME0 module production. (Left) Gas leak measurements for two ME0 modules, showing the exponential decrease of the over-pressure modeled as $P(t) = P_{0} e^{-t/\tau}$. The leak-rate parameter $\tau$ quantifies the pressure-decay timescale. The two modules shown have passed the test with measured values of $\tau = 4.36$~h and $11.14$~h and (Right) High-voltage divider test showing linearity between applied voltage and divider current with nominal and measured resistance values of the HV divider, respectively $R_{n} = 5.0~\mathrm{M\Omega}$ and $R_{m} = 4.95~\mathrm{M\Omega}$.}
\label{fig:ME0_QCresults}
\end{figure}

ME0 construction is distributed across multiple international sites: Panjab University, Delhi University, CERN (Geneva), Ghent University,
INFN, Università, and Politecnico di Bari, INFN (Frascati), RWTH Aachen University,
and Peking University. Components are first inspected and validated at CERN before being shipped to production centers for assembly and final testing. All quality control (QC) results are stored in the central GEM database ensuring full traceability. Figure~\ref{fig:ME0_QCresults} shows examples of the key validation tests performed during ME0 production.

 The QC chain ensures robust performance through several validation stages:
\begin{itemize}
    \item \textbf{Gas leak tests:} All modules exhibit leak rates corresponding to $\tau > 3$~h.
    \item \textbf{HV divider validation:} Resistance deviation remains below 3\% acceptance limit.
    \item \textbf{Gain and uniformity:} Effective gain $\sim 4.5\times10^{4}$, uniform across chambers.
    \item \textbf{Cosmic validation:} Efficiency and stability confirmed with cosmic stands.
\end{itemize}
About 50\% of the total production has been completed, with steady progress toward Long Shutdown 3 (LS3) installation ~\cite{DP126}.

\section{Performance Studies}
\subsection{Rate Capability @ GIF++}
The rate capability of ME0 chambers was evaluated during a dedicated test beam campaign at the CERN Gamma Irradiation Facility (GIF++) in July 2023~\cite{DP119}. A stack of four triple-GEM detectors equipped with the final VFAT3 front-end electronics was operated using an 80~GeV muon beam and a $^{137}$Cs $\gamma$ source. The detector was filled with an Ar/CO$_2$ (70:30) gas mixture, and events were triggered using two 30$\times$30~cm$^2$ scintillators.  

\begin{figure}[htbp]
\centering
\begin{minipage}[b]{0.45\textwidth}
\centering
\includegraphics[width=\textwidth]{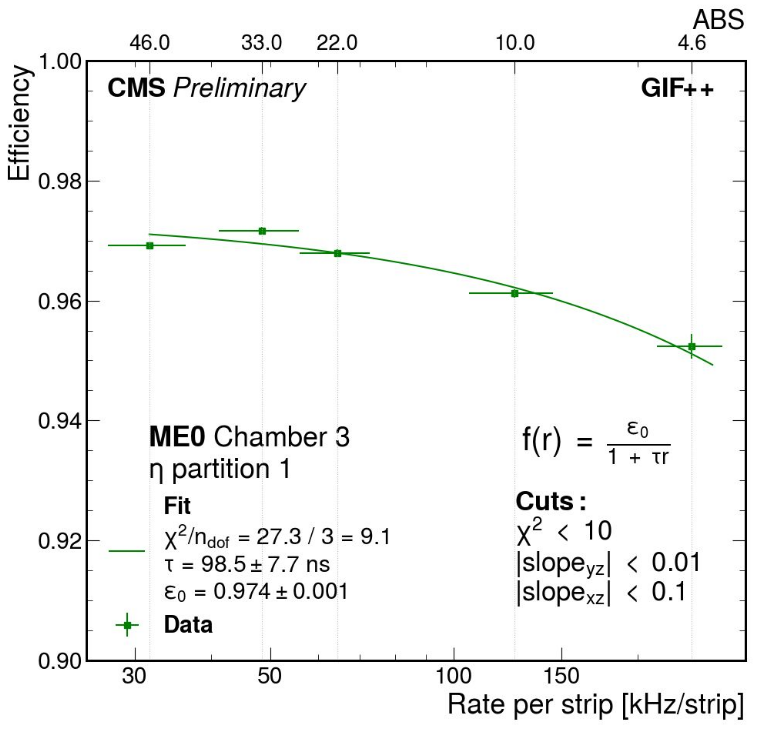}
\end{minipage}
\hfill
\begin{minipage}[b]{0.46\textwidth}
\centering
\includegraphics[width=\textwidth]{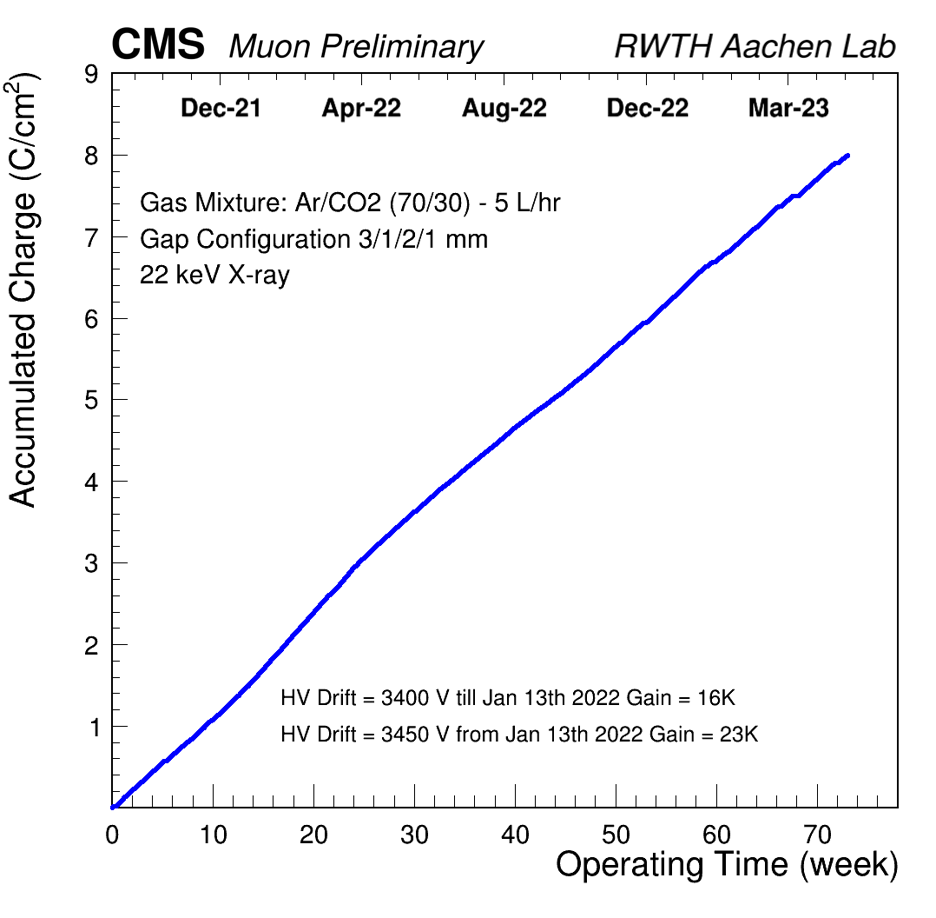}
\end{minipage}
\caption{ME0 performance results: (Left) Efficiency as a function of hit rate measured at GIF++, showing stable performance above 97\% with fitted $\tau$ between 98--182~ns~\cite{DP119} performed in the ABS (Aging and Beam Studies) setup and (Right)~Gain stability during aging measurements, where the integrated charge corresponds to the 
expected exposure in the hottest $\eta$ region~\cite{GEMPublic}.}
\label{fig:ME0_perf_summary}
\end{figure}

Muon efficiency was measured as a function of the background rate per strip, as shown in Figure~\ref{fig:ME0_perf_summary} (Left). There is a long efficiency plateau at $\sim$97\% up to several hundred kHz/cm$^2$, with characteristic dead time ($\tau$) values ranging between 98 and 182~ns, consistent with VFAT3 timing behavior. It was also observed that the cluster sizes for background hits are small, effectively reducing the chamber dead time. The efficiency and dead time measurements confirm that the ME0 design maintains stable performance under HL-LHC conditions.

\subsection{Longevity Studies}
Long-term irradiation tests demonstrate that the ME0 chambers can withstand the total accumulated charge expected over their full operational lifetime. During the aging campaign performed at RWTH Aachen lab, the chambers were exposed to an integrated charge of 8~C/cm$^{2}$, corresponding to the charge anticipated in the hottest $\eta$ region of the detector under HL-LHC conditions when applying a safety factor of unity, as shown in Fig.~\ref{fig:ME0_perf_summary} (Right). After correcting for temperature and pressure variations, the measured gas gain remained stable throughout the irradiation period, indicating no observable degradation in performance.

\subsection{Timing Resolution}
\begin{figure}[htbp]
\centering
\includegraphics[width=0.6\textwidth]{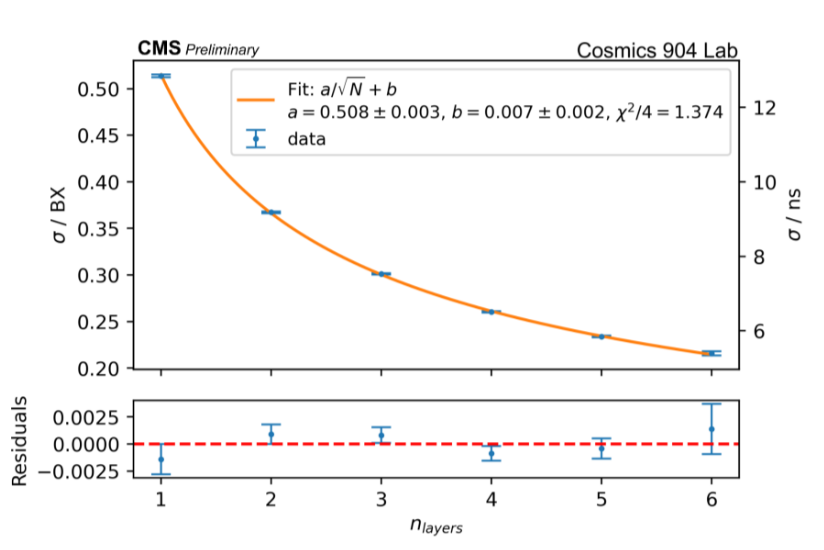}
\caption{Timing resolution of single ME0 layers and full stack measured with cosmic tests~\cite{DP089}.}
\label{fig:ME0_timeresolution}
\end{figure}

Timing performance was evaluated using cosmic muons at the CMS 904 laboratory and during test beams at GIF++. Single-layer chambers achieved a timing resolution of 10–12~ns, improving to $\sim$5.4~ns for the complete six-layer stack~\cite{DP089} as shown in Figure~\ref{fig:ME0_timeresolution}. This ensures correct bunch-crossing assignment and effective out-of-time background rejection.

At GIF++, the ME0 stack was further tested under simultaneous muon and $\gamma$ irradiation. The efficiency remained $\sim$95\% and stable across layers, with slight variations arising from geometrical shielding effects~\cite{DP019}. These results confirm that the ME0 satisfies HL-LHC timing requirements.

\section{Summary and Outlook}
The ME0 detector represents a key upgrade to the CMS muon system for the HL-LHC, extending acceptance to $|\eta| \approx 2.8$ and enabling efficient muon triggering and tracking in high-radiation regions. With validated QC procedures, stable performance at high rates, and demonstrated longevity, ME0 production and validation is proceeding toward full installation during LS3 (2027). 

\section*{Acknowledgments}
This work was supported by the Department of Science and Technology (DST) and  Anusandhan National Research Foundation (ANRF), Government of India.

\end{document}